\documentclass[11pt]{elsart}
\usepackage{amssymb}
\usepackage{amsmath}
\usepackage[dvips]{graphicx}
\input{tcilatex}
\journal{ }
\begin{document}

\begin{frontmatter}

\title{\large Interfaces endowed with non-constant surface energies revisited with the d'Alembert-Lagrange principle}

\author{Henri Gouin}
\ead{henri.gouin@univ-amu.fr; henri.gouin@yahoo.fr}

\address {  {
University of Aix-Marseille\   \     and\ \ M2P2, C.N.R.S.  U.M.R.  7340,\\  Marseille, France}  }%

\journal{\textbf{\ \ MEMOCS}, vol. \textbf{1}, n$^o$ 3  (2013).
\qquad\qquad\qquad\qquad\qquad\qquad\qquad\qquad\qquad\qquad \qquad\qquad\qquad\qquad\qquad \qquad\qquad\qquad\qquad\qquad }

\address{}

\begin{abstract}
{The equation of motions and
the
conditions on surfaces and edges between fluids and solids in presence of  non-constant \emph{surface energies}, as  in the case
of \emph{surfactants}
attached  to  the fluid particles at the interfaces, are revisited under the principle of virtual work. We point out
that
 adequate behaviors  of surface concentrations
may
drastically modify the \emph{surface tension} which naturally appears in the Laplace and the Young-Dupr\'e equations. Thus, the principle
of virtual work points out a strong  difference
between the
two revisited
concepts of surface energy and surface tension. }

\end{abstract}

\end{frontmatter}

\section{Introduction}
This paper develops the principle of virtual work due to d'Alembert-Lagrange \cite{Serrin} (\footnote{%
The principle of virtual  work  is also referred to in the
literature as the \emph{principle of virtual power} while virtual displacements are called   virtual velocities \cite{Germain, Germain1}.}) when different  phases of fluids are in contact through singular surfaces or interfaces. The study is first presented without constitutive assumption for surface energies but the displacement fields are considered for a simple material corresponding to the first-gradient theory. The d'Alembert-Lagrange principle  allows us to obtain   equation of motion and   boundary conditions of mechanical nature and is able to be extended to more complex materials with microstructures \cite{Maugin} or to multi-gradient theories \cite{Gouin3}.  Here, we aim to emphasize the formulation of the principle of virtual work when the interfaces are endowed with  non-constant surface energies:
the surfaces have their own material properties independent of the bulks and are embedded in the physical space, which is a three dimensional metric space. The surface energy density is taken into account and  naturally  comes into in the boundary conditions as the Laplace  and the Young-Dupr\'e  equations by using variations associated with the virtual displacement fields. To do so, it is necessary to propose a constitutive equation of the surface energy;    defining this is a main purpose of the paper. Such a presentation is similar to the one of the deformational and
configurational mechanics \cite{Steinmann};  the method is analogous with the one employed in \cite{Germain,Germain1,Maugin} but with powerful differential geometry  tools  as in \cite{Fosdick}. However, the mathematical tools are adapted to the linear functional of  virtual displacement fields and not to the integral balance laws over nonmaterial interfaces separating
fluid phases as in \cite{Cermelli}. \newline
Consequently, the main result  of the paper is to propose a general form of the linear functional with interfaces in first-gradient theory which points out the significance of constitutive behaviors for the surface energies and  highlights a   strong difference between the notions of surface energy and surface tension. Fischer \emph{et al} emphasized a thermodynamical definition of surface energy, surface tension and surface stress for which surface tension and surface stress are identical for fluids \cite{Fischer}. Our presentation is not the same: without any thermodynamical assumption, the difference between surface energy and surface tension is a natural  consequence of the virtual work functional and the d'Alembert-Lagrange principle. The surface energy allows to obtain the total energy of the interfaces and the surface tension is directly generated  from the boundary conditions of the continuous medium.\newline In the simplest cases the two notions of surface energy and surface tension are  mingled, but it is not generally the case when the surface energy is non-constant along the interfaces.
To prove this property, we first focus on the simplest case of Laplace's capillarity and we obtain the well-known equations on   interfaces and  contact lines.\newline
Surfaces endowed with surface matter as in  the case of surfactants is a more complex case.
The last decades have seen the extension of surfactant applications in
many fields including biology and medicine \cite{Rosen}; surfactants can also be expected to play a major mechanical role in fluid and
solid domains. The versatility of surfactant
mainly depends on its concentration at interfaces.
It experimentally appears that surfactant or
surface-active agent is a substance present in liquids at very low
concentration rate and, when surface mass concentration is below the critical
micelle concentration, it is mainly absorbed onto interfaces and alters only the
interfacial free energies \cite{Gennes}. The interfacial free
energy per unit area (generally called  \emph{surface energy}) is the minimum
amount of work required to create an interface at a given temperature \cite%
{Edwards,slaterry} and the fact that surfactants can affect the mechanical
behaviors of interfaces must be modelized in order to predict and control
the properties of complete systems. \newline
In fact, \emph{our aim is not to study  the  general case of surfactants} proposed in the literature  but to focus on
the virtual work method  to
prove that simple behaviors of the surface energy depending on the mass
concentration  can drastically change  the capillary effects. So, the  concept of surface tension naturally appears
in the equations on surfaces and on lines. In this paper, we call \emph{surfactant} the matter distributed  only on the interfaces:
we consider the special case when surfactant molecules are insoluble in the
liquid bulk (the surface mass concentration is below the critical
micelle concentration \cite{Rosen}) and are attached  to fluid particles along the interfaces (without surface diffusion as in \cite{McBride}).

The manuscript is organized as follows:\newline
   Section 2  briefly  reminds some  results formally  presenting the principle of virtual work in its more general form by using the kinematics of a continuous medium and the notion of virtual displacement. The simplest example of the  Laplace model of capillarity concludes the Section.\newline
Section 3 deals with the case
when the interfaces are endowed with non-constant surface energy, whereby we essentially focus  on liquid in contact with  solid and   gas. The special case of surfactants as interface matter attached to the fluid particles is considered. The surface energy depends on the surface matter concentration. Such a property  drastically changes the boundary conditions on the interface by using   surface tension instead of   surface energy.\newline Section 4 deals with an explicit comparison between  surface energy and   surface tension only within deformational mechanics.\newline Section 5 is the conclusion in which some general extension can be forecasted.\newline The main mathematics tools are collected in a large appendix so that the presentation of the text is not cluttered with tedious calculations. The main mathematical tool is Relation (\ref{works}) which can be extended to more complex media.
\section{The virtual work for continuous medium}
In continuum mechanics, motions can be equivalently studied with either the Newton model of \emph{system of forces} or   the Lagrange model of \textit{%
the work of forces} \cite{Germain,Germain1}. The Lagrange model
does not derive from a variational approach but, at
equilibrium, the minimization of the energy
coincides with the zero value of a linear functional. Generally, the linear functional expressing the work of forces is related
to the theory of distributions; a decomposition theorem associated with
displacements (as $C^{\infty }$-test functions whose supports are compact
manifolds) uniquely determines a canonical zero order form (\textit{%
separated form}) with respect both to the test functions and the transverse
derivatives of the contact test functions \cite{Schwartz}.
In the same way that the Newton   principle is useless when we do not have any constitutive
equation for the system of forces, the d'Alembert-Lagrange principle is
useless when we do not have any constitutive assumption for the virtual work
functional.
\newline
The equation of motion and boundary conditions of a continuous medium derives from the
\textit{d'Alembert-Lagrange principle of virtual work}, which is an extension
of the same principle in mechanics of systems with a finite number of degrees of
freedom: \textit{\ For any virtual displacement, the motion is such that the
virtual work of forces is equal to the virtual work of mass accelerations}
\cite{Gouin3}.

\subsection{\textbf{The background of the principle of virtual work}}

The motion of a continuous medium is classically represented by a continuous
transformation $\mathbf{\varphi }$ of a three-dimensional space into the
physical set. In order to describe the transformation analytically, the
variables $\mathbf{X} = (X^{1},X^{2},X^{3})$ which single out individual
particles correspond to material or Lagrange variables; the variables $%
\mathbf{x}= (x^{1},x^{2},x^{3})$ corresponds to Euler variables. The
transformation representing the motion of a continuous medium is of the form
\begin{equation}
\mathbf{x=\varphi }\left( \mathbf{X,}t\right) \text{ \quad or \quad }x^{i}=\varphi
^{i}(X^{1},X^{2},X^{3},t)\, , \ i \in \left\{1, 2, 3\right\},  \label{motion}
\end{equation}
where $t$ denotes the time. At a fixed time  the transformation possesses an
inverse and continuous derivatives up to the second order except on singular
surfaces, curves or points. Then, the diffeomorphism $\mathbf{%
\varphi }$ from the set $D_{0}$ of the particle references into the physical
set $D$ is an element of a functional space $\mathbf{\wp}$ of the positions
of the continuous medium considered as a manifold with an infinite number of
dimensions.\newline
To formulate the d'Alembert-Lagrange principle of virtual work in continuum
mechanics, we remind the notion of \textit{virtual displacements}.
This notion is obtained by letting the displacements arise from variations
in the paths of particles. Let a one-parameter family of varied paths or
\emph{virtual motions} denoted by $\{\mathbf{\varphi }_{\eta }\}$, and
possessing continuous partial derivatives up to the second order, be
analytically expressed by the transformation
\begin{equation}
\mathbf{x=\Phi }\left( \mathbf{X,}t;\eta \right)  \label{vitual
motion}
\end{equation}
with $\eta \in O,$ where $O$ is an open real set containing $0$ and such
that $\mathbf{\Phi }\left( \mathbf{X,}t;0\right) =\mathbf{\varphi }\left(
\mathbf{X,}t\right) $ (the real motion of the continuous medium is obtained
when $\eta =0$). The derivative with respect to $\eta $ at $\eta =0$ is
denoted by $\delta $. In the literature, derivative $\delta $ is named
\textit{variation} and the \textit{virtual displacement} is the variation of
the position of the medium \cite{Serrin}\,. The virtual displacement is a
tangent vector to $\mathbf{\wp }$, functional space of positions, at $\mathbf{\varphi }$ ($\delta \mathbf{%
\varphi }\in T_{\mathbf{\varphi }}(\mathbf{\wp }))$. \ In the physical
space, the \textit{virtual displacement} $\delta \mathbf{\varphi }$ is
determined by the variation of each particle: the \textit{virtual
displacement} $\mathbf{\zeta}$ \textit{of the particle} $\mathbf{x}$ is such
that $\mathbf{\zeta}=\delta \mathbf{x}\ $ when at $\eta =0,\ \left\{\delta\mathbf{X}=0, %
\delta t =0,  \delta \eta =1\right\}$  and we associate the field of
tangent vectors to $D$:
\begin{equation*}
{\mathbf{x}}\in D\ \mathbf{\rightarrow \mathbf{\zeta} }=\mathbf{\psi (x)}%
\equiv \frac{\partial \mathbf{\Phi }}{\partial \eta }\left| _{\eta
=0}\right. \in T_{\mathbf{x}}(D),
\end{equation*}
where $T_{\mathbf{x}}(D)$ is the tangent vector bundle to $D$ at $\mathbf{x}$ 
(Figure 1).\newline

\begin{figure}[h]
\begin{center}
\includegraphics[width=9cm]{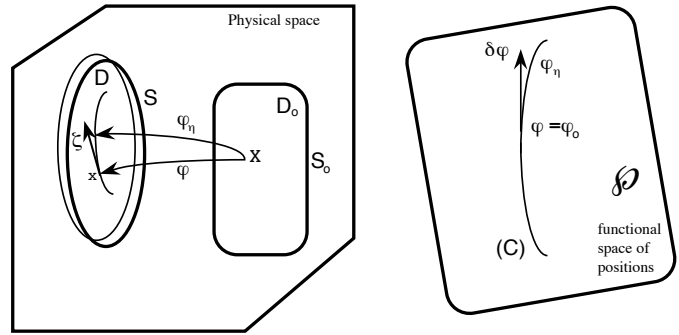}
\end{center}
\caption{The boundary $S$ of $D$ is represented by a thick curve and its
variation by a thin curve. The variation $\protect\delta \mathbf{\protect%
\varphi} $ of the family $\{\mathbf{\protect\varphi }_\protect\eta \}$ of
varied paths belongs to $T_{\mathbf{\protect\varphi }}(\mathbf{\wp })$,
tangent space of $\mathbf{\wp }$ at $\mathbf{\protect\varphi }$.}
\label{fig1}
\end{figure}

The  virtual work  concept, dual of Newton's method, can be written in the following form:\newline
\textit{The virtual work $\delta\tau$   is a linear functional
value of the virtual displacement,}
\begin{equation}
\delta \tau =<\Im ,\delta \mathbf{\varphi }>  \label{virtual work of
forces}
\end{equation}
\textit{where} $< .\, , . >$ \textit{denotes the inner product of} $\Im $
\textit{and} $\delta \mathbf{\varphi }$, \textit{with} $\Im $ \textit{%
belonging to the cotangent space} $T_{\mathbf{\varphi }}^*(\mathbf{\wp })$
\textit{of} $\mathbf{\wp }$ \textit{at} $\mathbf{\varphi }$.

In Relation (\ref{virtual work of forces}), the medium in position $\mathbf{%
\varphi }$ is submitted to  covector $\Im $ denoting all the "stresses" in mechanics. In
the case of motion, we must add  the inertial forces, corresponding to the
accelerations of masses, to the volume forces.

\textit{The d'Alembert-Lagrange principle of virtual work is expressed as
follows}

\textit{\centerline {\textbf{For all virtual displacements, the virtual work is null}.}%
}

\noindent The principle leads
to the analytic representation
\begin{equation*}
\forall \ \delta \mathbf{\varphi }\in \mathit{T}_{\mathbf{\varphi }}(\mathbf{%
\wp }),\ \delta \tau =0  \label{d'Alembert-Lagrange}
\end{equation*}
\textbf{Theorem}: \,  \textit{If expression (\ref{virtual work of forces}) is a
distribution expressed in a separated form \cite{Schwartz}, the d'Alembert-Lagrange
principle yields the equation of motion and boundary conditions in the form%
}\ \ $\Im =0 $\,.

The virtual displacement is submitted to constraints coming from the
constitutive equations and geometrical assumptions such as the mass conservation. Consequently, the
constraints are not expressed by Lagrange multipliers but are directly taken
into account by the variations of the constitutive equations. The equation
of motion and boundary conditions result from the explicit expression of $%
\delta \tau$ associated with the considered physical problem. As a first example,  the simplest case of
 theory of capillarity at equilibrium is considered.

\subsection{\textbf{The classical Laplace theory of capillarity}}

Liquid-vapor and two-phase interfaces are represented by  material surfaces
endowed with an energy related to the Laplace free energy of capillarity. When working far from
critical conditions, the capillary layer has a thickness equivalent to a few
molecular beams \cite{Ono,domb}   and the interface
appears as a geometrical surface separating two media, with its own characteristic
behavior and energy properties \cite{Levitch}. The domain $D$ of
a compressible fluid (liquid) is immersed in a three Euclidian space. The boundary of the   domain $D$  is a surface $S$ shared in  $N$ parts $%
S_{p}$ of class $C^{2}$, $(p=1,...,N)$ (Fig. 2). We denote by $(R_{m})^{-1}$
the mean curvature of $S$;  the union of the limit edges $\Gamma _{pq}$ between
surfaces $S_{p}$ and $S_{q}$ is assumed to be of class $C^{2}$ and $\mathbf{t%
}_p $ is the tangent vector to $\Gamma_{p}= \bigcup\, \Gamma _{pq},  q =1,...,N \ {\rm with}\ q\neq p $, oriented by the unit external
vector to $D$ denoted $\mathbf{n}_p$; $\mathbf{n}_p^\prime = \mathbf{t}%
_p\times \mathbf{n}_p$ is the unit external normal vector to $\Gamma _{p}$
in the tangent plane to $S_{p}$; the edge $\Gamma$ of $S$ is the union of the edges $%
\Gamma _{p}$ of $S_{p}$.
\begin{figure}[h]
\begin{center}
\includegraphics[width=11cm]{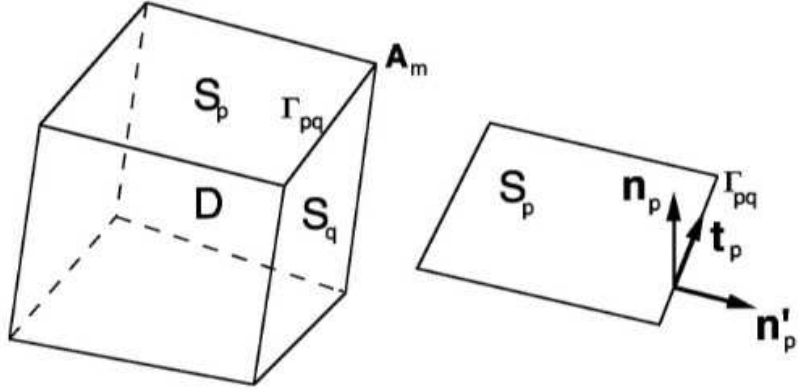}
\end{center}
\caption{The set $D$ has a surface boundary $S$ divided in several parts.
The edge of $S$ is denoted by $\Gamma $ which is also divided into several
parts with end points $\mathbf{A}_{m}$. }
\label{fig2}
\end{figure}
\newline
To first verify the well-founded of the model, we consider the explicit expression of the functional $\delta \tau $ for
compressible fluids with capillarity in non-dissipative case.
The variation of the total  energy $E$ of such a  fluid results from
the variation of the sum of the local density of energy  integrated on
the domain $D$ and the variation of the local density of surface energy integrated on its boundary $S$;
to these variations, we must add the work of volume force $\rho\, \textbf{f}$ in $D$,
surface force $\mathbf{T}$ on $S$ and    line force $\mathbf{L}$ on $%
\Gamma $. Such an amount represents,  for the domain $D$, the virtual work of forces of the
compressible fluid with capillarity.\newline
\noindent The Laplace theory of capillarity introduces the notion of
surface energy (or superficial energy) on surfaces such that, for a compressible liquid with
capillary effects  on the wall boundaries, the total  energy of
the fluid writes in the form
\begin{equation*}
E=\int \int \int_{D}\rho \ \alpha (\rho )\ dv+\int \int_{S}\sigma \ ds,\quad
\mathrm{with}\quad \int \int_{S}\sigma \ ds\equiv \sum_{p=1}^{N}\int
\int_{S_{p}}\sigma _{p}\ ds,
\end{equation*}%
where $\rho $ is the matter
density, $\alpha (\rho )$ is the fluid specific energy ($\rho\,\alpha (\rho )$ is the volume energy)  and the   coefficients $\sigma _{p}$ are the surface
energy densities on each surface $S_{p}$ represented -for the sake of simplicity- by
$\sigma $ on $S$ (\footnote{Our aim is not to consider the thermodynamics of interfaces. Consequently, $\alpha$ and $\sigma$ are not considered as functions of thermodynamical variables such as   temperature or   entropy.}). Surface integrations are associated with the metric
space. As proved in Appendix, the variation of the deformation gradient
tensor $F=\partial \mathbf{x}/\partial \mathbf{X}$ (with components $\{\partial
x_{i}/\partial X_{j}\}$) of the mapping $\mathbf{\varphi}$ combined with the
mass conservation and the variation of $\sigma $ allow to obtain the
variation $-\delta E$ (see Eq. (\ref{varsurf0}) in Appendix); then the
independent variables come from  the position $\mathbf{x}$ of the
continuous medium.\newline
The virtual work of volume forces defined on $D$ is generally in the form
\begin{equation*}
\int \int \int_{D}\rho \ \mathbf{f}^{T}\mathbf{\zeta }~dv,\quad \mathrm{with}%
\quad \mathbf{f}=-\func{grad}\ U  \label{volume forces}
\end{equation*}%
where $U(\mathbf{x})$ is a potential per unit mass and superscript $^T$ denotes
the transposition. The virtual work of surface and line forces defined on $S$
and   $\Gamma$ are respectively,
\begin{equation*}
\int \int_{S} \ \mathbf{T}^{T}\mathbf{\zeta }~ds\quad \mathrm{and}\quad
\int_{\Gamma } \ \mathbf{L}^{T}\mathbf{\zeta }~dl.  \label{surfaceforces}
\end{equation*}%
Consequently, the total virtual work of forces $\delta\tau$ is
\begin{equation*}
\delta\tau=-\delta E +\int \int \int_{D}\rho \ \mathbf{f}^{T}\mathbf{\zeta }%
~dv+\int \int_{S} \ \mathbf{T}^{T}\mathbf{\zeta }~ds+\int_{\Gamma } \
\mathbf{L}^{T}\mathbf{\zeta }~dl.
\end{equation*}
From Eqs. (\ref{varsurf0}) and (\ref{pressure}) in Appendix, we obtain
\begin{eqnarray}
\delta \tau &\equiv &\int \int \int_{D}\left( -\mathrm{{grad}^{T}}p+\rho\,
\mathbf{f}^{T}\right) \mathbf{\zeta }~dv\ + \int_{\Gamma }\left( \mathbf{L}%
^{T}-\sigma \,\mathbf{n}^{\prime T}\right) \mathbf{\zeta }~dl\
\label{Laplace} \\
&& + \int \int_{S}\left[ -\delta \sigma +\left\{ \left({p}\, +\frac{2\,\sigma }{R_{m}}\right)\,%
\mathbf{n}^{T}+\mathrm{grad}^{T}\sigma \left( \mathbf{1}-\mathbf{nn}%
^{T}\right) + \mathbf{T}^{T}\right\} \mathbf{\zeta }%
\right] ~ds  \notag
\end{eqnarray}%
where $p\equiv \rho ^{2}\alpha ^{\prime }(\rho )$ is the pressure of the liquid
\cite{Rocard}, $\delta\sigma$ denotes the variation of the surface
energy $\sigma$ and   $\textbf{1}$ denotes the identity tensor. When $\sigma$
is constant we get $\delta\sigma = 0$; then,
\begin{eqnarray*}
\delta \tau &\equiv &\int \int \int_{D}\left( -\mathrm{{grad}^{T}}p+\rho\,
\mathbf{f}^{T}\right) \mathbf{\zeta }~dv\ + \int \int_{S}  \left\{ \left(p+\frac{%
2\,\sigma }{R_{m}}\right)\,\mathbf{n}^{T} +   \mathbf{T}%
^{T}\right\} \mathbf{\zeta }  ~ds \\
&& + \int_{\Gamma }\left( \mathbf{L}^{T}-\sigma \,\mathbf{n}^{\prime
T}\right) \mathbf{\zeta }~dl.
\end{eqnarray*}
and the d'Alembert-Lagrange principle yields the equation of equilibrium on $%
D$,
\begin{equation}
-p_{,i}+\rho f_{i}=0\quad \mathrm{or}\quad -\mathrm{grad}\,p+\rho \,\mathbf{f%
}=0 .  \label{equilibrium1}
\end{equation}%
The condition on boundary surface $S$ is,
\begin{equation}
\left(p+ \dfrac{2 \, \sigma}{R_{m}}\right)\, n_{i}+T_{i}=0\quad \ \mathrm{or}\quad
\ \left(p\,+\dfrac{2\,\sigma }{R_{m}}\,\right)  \mathbf{n}+\mathbf{T} =0\ ,
\label{surface1}
\end{equation}
where, for an external fluid bordering $D$, $\mathbf{T}= -P\,\mathbf{n}$, with $P$   value of
the pressure in the external fluid. On the lines, it is necessary to
consider the partition of $S$ such that the edge $\Gamma _{pq}$ is common to
$S_p$ and $S_q$,
\begin{equation}
\sigma _{p}\,n_{pi}^{\prime }+\sigma _{q}\,n_{qi}^{\prime }-L_{i}=0\qquad \
\mathrm{or}\qquad \ \sigma _{p}\,\mathbf{n}_{p}^{\prime }+\sigma _{q}\,%
\mathbf{n}_{q}^{\prime }-\mathbf{L}=0.  \label{edge1}
\end{equation}%
Surface condition (\ref{surface1}) is the \textit{Laplace equation} and line condition (\ref{edge1}) is the
\textit{Young-Dupr\'{e} equation} with a line tension $\textbf{L}$.\newline
It is interesting to note that in  \cite{Steigmann1}, Steigmann and Li used the principle of virtual work by utilizing a system of line coordinates on boundary surfaces and lines. By introducing the free energy per unit area of interfaces and the free energy per unit of contact curve, they obtained Laplace's equation and a generalization of Young-Dupr\'e's equation of equilibrium; moreover, by employing necessary conditions for energy-minimizing states of fluid systems they got a demonstration that the line tension associated with a three-phase contact curve must be nonnegative.\newline

When $\sigma$ is not constant but $\delta\sigma=0$, we obtain
the same equations for Eq. (\ref{equilibrium1}) and Eq. (\ref{edge1}) but
Eq. (\ref{surface1}) on $S$ is replaced by
\begin{equation*}
\left(p\, +\dfrac{2\,\sigma }{R_{m}}\right)  \mathbf{n} + \left(
\mathbf{1}-\mathbf{nn}^{T}\right)\mathrm{grad}\,\sigma +\mathbf{T}=0 .  \label{surface2}
\end{equation*}
 The additive term $\left( \mathbf{1}-\mathbf{nn}^{T}\right)\mathrm{grad}%
\,\sigma = \mathrm{grad}_{tg}\sigma$ is the tangential part  of $%
\mathrm{grad}\,\sigma$ to the surface $S$. This term corresponds to a shear stress necessarily
balanced by the tangential component of $\mathbf{T}$. Such is the case
 when $\sigma$ is defined on $S_0$ image of $S$ in
the reference space $D_0$ (then, $\sigma = \sigma_0(\mathbf{X})$). We understand
the importance  of the surface energy constitutive behavior; this questioning is emphasized in the following section.

\section{Capillarity of liquid in contact with solid and gas in presence of
non-constant surface energy}
We have seen in the previous section that the problem associated with the behavior of the surface energy is the key point to obtain the boundary conditions on interfaces and contact lines bordering the fluid bulk. In this section we consider a very special case of surfactant: the interfaces are endowed with a  concentration of matter which affects the surface energy. The surface matter is attached to the particles of the fluid such that they obey together to the same Eq. (\ref{motion}) of motions and Eq. (\ref{vitual motion}) of  virtual motions. We  consider a more general case than in Section 2.2: we study the motion of the continuous medium with viscous forces. This viscosity affects not only the equation of motion but also the boundary conditions.
\subsection{\textbf{Geometrical description of the continuous medium}}
A drop of liquid fills the set $D$ and lies on the surface of a solid. The
liquid drop is also bordered by a gas. All the interfaces between liquid,
solid and gas are assumed to be regular surfaces.
\begin{figure}[h]
\begin{center}
\includegraphics[width=10cm]{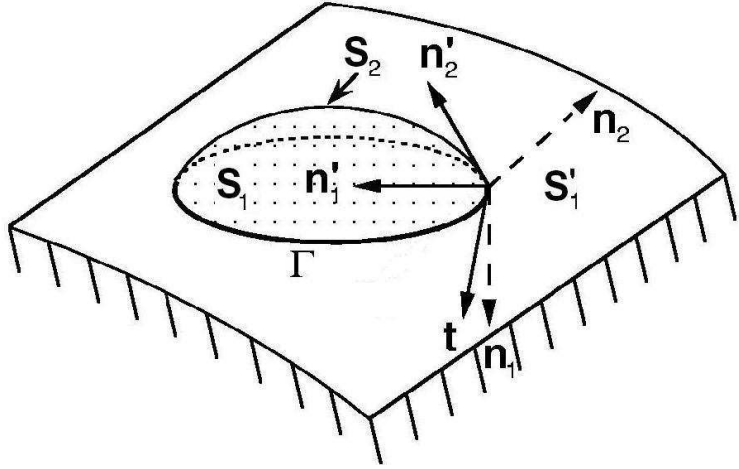}
\end{center}
\caption{ A liquid in drop form
lies on a solid surface. The liquid is bordered with a gas and a solid; $S{%
_1}$ is the boundary between liquid and solid; $S^{\prime}_{1}$ is the
boundary between gas and solid; $S_{2}$ is the interface between liquid and
gas ; $\mathbf{n}_1$ and $\mathbf{n}_2$ are the unit normal vectors to $%
S_{1} $ and $S_{2}$, external to the domain of liquid; the edge $\Gamma$ (or contact line) is common to $S_{1}$ and $%
S_{2}$ and $\mathbf{t}$ is the unit tangent vector to $\Gamma$ relative to $%
\mathbf{n}_1$; $\mathbf{n^{\prime}}_1 = \mathbf{n}_{1}\times \mathbf{t}$ and $\mathbf{n^{\prime}}_2=\mathbf{n}_{2}\times (-\mathbf{t})$ are the
binormals to $\Gamma$ relative to $S_{1}$ and $S_{2}$, respectively. }
\label{fig3}
\end{figure}
We call $\sigma_{_{S _{1}}}$ and $\sigma_{_{S _{2}}}$ the values of the
surface energies of $S _{1}$ and $S _{2}$, respectively (see Fig. \ref%
{fig3}). These energies may depend on each point of the boundary of $D$.
Afterwards, on the domain $S _{1}^{\prime }$, the surface energy between
gas and solid is neglected \cite{Adamson}. The liquid drop is submitted to
a volume force $\rho\, \mathbf{f}$. The $\mathit{external}$ surface force
on $D$ is modelized with two constraint vector fields $\mathbf{T}_{1}$ on
the solid surface, $S _{1} $ and $\mathbf{T}_{2}$ on the free surface, $S _{2}$%
. The line tension $\mathbf{L}$ is assumed to  be null.\newline
By using the principle of virtual work, we aim to write the motion
equation of the liquid drop  and the conditions on surfaces and line
bordering the liquid drop.

\subsection{\textbf{Surfactant attached to interfacial fluid particles}}
To express the behavior of the surface energy, we need to represent first the equation of the surface matter density.\newline
By using the mapping $\mathbf{\varphi}$, the  set $D_{0} $ of boundary $%
S_{0}$ has the image $D $ of boundary $S$. We assume there exists an
insoluble surfactant with a surface mass concentration $c_{0}$ defined on $%
D_{0}$ of image $c$ in $D$ \cite{Levitch,Adamson,Defay}. Let us consider
the case when \textit{the surfactant is attached to the fluid particles on
the surface }$S$,\ i.e.
\begin{equation}
c_{0}=c_{0}(\mathbf{X}),\quad \mathrm{{where}\ \mathbf{X} \in S_{0}\,.}
\label{concentration1}
\end{equation}%
The mass conservation  of the surfactant on the surface $S$ requires that for
any subset $S_{0}^{\ast }$ of $S_{0}$, of image $S^{\ast }$ subset of $S$,%
\begin{equation}
\int \int_{S^{\ast }}c\ ds=\int \int_{S_{0}^{\ast }}c_{0}\ ds_{0}.
\label{surfactint}
\end{equation}%
Relation (\ref{surfactint}) implies
\begin{equation}
c\ \det F\ \mathbf{n}_{0}^{T}F^{-1}\mathbf{n}=c_{0}\quad \mathrm{with} \quad
\mathbf{n}_{0}^{T}=\frac{\mathbf{n}^{T}F}{\sqrt{(\mathbf{n}^{T}FF^{T}\mathbf{%
n})}} \,,  \label{concentration2}
\end{equation}
where $\mathbf{n}_{0}$ denotes the unit normal vector to $S _{0}$. The proof
of Rel. (\ref{concentration2}) is given in Appendix.\newline
From Rel. (\ref{concentration1}) and Rel. (\ref{concentration2}), we obtain: \newline
Firstly, the conservation of the surface concentration of the surfactant,
\begin{equation}
\frac{dc}{dt}+c\left( \func{div}\mathbf{u}-\mathbf{n}^{T}\mathbf{D}\ \mathbf{%
n}\right) =0,  \label{conservation}
\end{equation}%
where $\mathbf{u}$ is the fluid velocity vector and $\mathbf{D}= \frac{1}{2}\left(
{\partial\mathbf{u}}/{\partial \mathbf{x}}+\left( {\partial \mathbf{u}}/{%
\partial \mathbf{x}}\right) ^{T}\right) $ denotes the rate of the
deformation tensor of the fluid. The term\thinspace\ $\func{div}\mathbf{u}-%
\mathbf{n}^{T}\mathbf{D}\ \mathbf{n}$ \thinspace expresses the tangential
divergence relative to the surface $S$. \newline
Secondly, the variation of the mass concentration of the surfactant,
\begin{equation}
\delta c+c\left[ \func{div}\mathbf{\zeta}-\mathbf{n}^{T}\frac{\partial
\mathbf{\zeta}}{\partial \mathbf{x}}\,\mathbf{n}\right] =0.
\label{varsurfactant}
\end{equation}
The proofs of Relations (\ref{conservation}) and (\ref{varsurfactant}) are also
given  in Appendix. In the case when the surface energy $\sigma$ is a
function of the surfactant concentration,
\begin{equation*}
\sigma=\sigma(c),  \label{energy and concentration}
\end{equation*}
we deduce $\delta \sigma=\sigma^{\prime }(c)\,\delta c$.
If we denote
\begin{equation}
\gamma =\sigma-c\ \sigma^{\prime }(c)\,,  \label{legendre}
\end{equation}
which is the Legendre transformation of $\sigma$ with respect to $c$, then by taking
Rel. (\ref{varsurf0})   into account, we obtain in Appendix,
\begin{equation}
\delta E=-\int \int_{S}\left[ \, \frac{2\gamma }{R_{m}}\,\mathbf{n}^{T}+%
\mathrm{{grad}^{T}\gamma \ \left( \mathbf{1}-\mathbf{nn}^{T}\right) }
\right]\mathbf{\zeta} ~ds+\int_{\Gamma}\gamma \ \mathbf{n}^{\prime T}\mathbf{%
\zeta}~dl.  \label{variationE}
\end{equation}
As we shall see in Section 4, $\gamma$ is the \emph{surface tension} of the interface $S$.
The variation $\delta c$ of the concentration has an important consequence
on the surfactant behavior  and  the surfactant behavior is essential to
determine the virtual work of the liquid drop.\newline
Relation (\ref{legendre}) can easily be extended to several surfactants:  if $%
\sigma = \sigma (c_{1},\ldots,c_{n})$ where $\displaystyle c_{i},\
i\in\{1,\ldots,n\}$ are the concentrations of $n$ surfactants, then
\begin{equation*}
\gamma =\sigma -\sum_{i=1}^{n}c_{i}\,\frac{\partial \sigma }{\partial c_{i}}%
\ ,
\end{equation*}
corresponding to the Legendre transformation of $\sigma$ with respect to $%
c_{i}, \ i\in\{1,\ldots,n\}$ and Eq. (\ref{variationE}) is always valid.

\subsection{\textbf{Governing equation of motion and boundary conditions}}
As previously indicated, we do not consider the thermodynamical problem of interfaces, but for example, when the medium is isothermal, $\alpha$ can be considered as the specific free energy of the bulk and $\sigma$ the free surface energy of the interface.\newline
The use of virtual displacements yields a linear functional of virtual
work, sum of several partial works. To enumerate the works of forces, we
have to consider how they are obtained in the literature \cite%
{Germain,Germain1,Gouin3}. The virtual work expressions of volume force $\rho\,\mathbf{f}$,
surface force $\mathbf{T}$ and liquid pressure $p$ are the same as in Section
2.2.

a) For fluid motions, the virtual work of mass impulsions is
\begin{equation*}
- \int \int \int_{D} \rho \ \mathbf{a }^{T}\mathbf{\zeta}~dv,
\end{equation*}
where $\mathbf{a}$ is the acceleration vector. \newline
b) For dissipative motions, we must add
 the virtual work of viscous stresses
\begin{equation*}
-\int \int \int_{D}\mathtt{tr}\left( \mathbf{\tau}_{v}\frac {\partial \mathbf{\zeta%
}}{\partial \mathbf{x}}\right) ~dv,
\end{equation*}
where $\mathbf{\tau}_{v}$ denotes the viscous stress tensor usually written
in the form of Navier-Stokes \cite{Adamson}. Taking account of the relation
\begin{equation*}
\mathtt{tr} \left(\mathbf{\tau}_{v}\, \frac{\partial\mathbf{\zeta}}{\partial%
\mathbf{x}}\right) = \func{div}(\mathbf{\tau}_{v}\, \mathbf{\zeta})-({\func{%
div}}\mathbf{\tau}_{v})\,\mathbf{\zeta} , \label{viscous}
\end{equation*}
an integration by parts using Stokes' formula in Eq. (\ref{variationE}) for
the virtual work of interfacial forces,  and  relations $\mathbf{n^{\prime}}_1 =  \mathbf{n}_{1}\times \mathbf{t}$, $\mathbf{n^{\prime}}_2=-\mathbf{n}_{2}\times \mathbf{t}$, allow to obtain the virtual work of
forces applied to the domain $D$
\begin{equation}
\begin{array}{l}
\delta\tau = \displaystyle \int \int \int_{D} \left( -\,{\func{grad}}^{T}p  +
\func{div}\mathbf{\tau}_{v} + \rho \,\mathbf{f}^{T} - \rho \ \mathbf{a }%
^{T}\right)\mathbf{\zeta}~dv \\
\ +\displaystyle \int \int_{S _{1}}\left[ {{\func{grad}}^{T} \gamma
_{1}\left(\mathbf{1}-\mathbf{n}_{1}\mathbf{n}_{1}^{T}\right) +\left( p+{\frac{2\gamma _{1}}{R_{m_{1}}}}\right)\,\mathbf{n}%
_{1}^{T}} -{\mathbf{n}_{1}^{T}}\mathbf{\tau}_{v}+\mathbf{T}_{1}^{T}\right] \mathbf{\zeta}~ds \\
+\displaystyle \int \int_{S _{2}}\left[  {{\func{grad}}^{T} \gamma
_{2} \left(\mathbf{1}-\mathbf{n}_{2}\mathbf{n}_{2}^{T}\right) +\left( p+{\frac{2\gamma _{2}}{R_{m_{2}}}} \right)\, \mathbf{n}%
_{2}^{T}} -{\mathbf{n}_{2}^{T}}\mathbf{\tau}_{v}+\mathbf{T}_{2}^{T}\right] \mathbf{\zeta}~ds \\
\quad\displaystyle +\int_{\Gamma}\left( \,\gamma _{1}\mathbf{n}_{1}^{^{\prime }T} -
\gamma _{2}\mathbf{n}_{2}^{^{\prime }T}\right)\mathbf{\zeta}~dl,%
\end{array}
\label{works}
\end{equation}
where $R_{m_{i}}$ denotes the mean radius of curvature of $S_i$,  $\gamma_i$ denotes the surface tension of $S_i$ and $%
\mathbf{T}_{i}$ the surface force on $S_i$, $i \in \{1,2\}$; $\mathbf{T}_{2}
=  -\,P \mathbf{n}_{2}$, where $P$ is the pressure in the external gas to the domain $D$.\newline
The field of virtual displacement $\mathbf{x\in \ }D \longrightarrow \mathbf{%
\zeta(x)\in \ }T_{\mathbf{x}}(D)$ must be tangent to the solid (rigid) surface $S_1$%
. The fundamental lemma of variational calculus yields  the equation
of motion associated to   domain $D$, the conditions on surfaces $S_1$, $S_2$
and the condition on  contact line $\Gamma$.\newline
Due to the fact that Eq. (\ref{works}) is expressed in separate form in the sense of
distributions \cite{Schwartz}, the d'Alembert-Lagrange principle implies
that $\forall\ \mathbf{\zeta(x)\in \ }T_{\mathbf{x}}(D)$ tangent to $S_1 $, each of
the four integrals of Eq. (\ref{works}) is  null. Then, we obtain
equations on $D$, $S_1$, $S_2$ and $\Gamma$, respectively.

$\bullet \quad $ We get \textit{the equation of motion} in $D$
\begin{equation}
\rho \ \mathbf{a }+ \func{grad}%
p =  \left(\func{div}\mathbf{\tau}_{v}\right)^T + \rho \,\mathbf{f}.  \label{NS}
\end{equation}
Equation (\ref{NS}) is the Navier-Stokes equation for compressible fluids
when $\mathbf{\tau}_{v}$ is written in the classical linear form by using the rate
of the fluid deformation tensor, $\mathbf{\tau} _{v}= \lambda  (\mathrm{{tr%
}\, \mathbf{D)}\,\mathbf{1}+ 2\mu\, \mathbf{D}}$. We
may add a classical condition for the velocity on the boundary as the adherence condition.

$\bullet \quad $ We get \textit{the condition} on surface $S _{1}$

The virtual displacement is tangent to $S _{1}$; the constraint $\mathbf{n}%
_{1}^{T} \mathbf{\zeta}=0$ implies there exists a scalar Lagrange multiplier
$\mathbf{x\in }$ $S _{1}\longrightarrow \chi \left( \mathbf{x}\right) \in \Re
$, such that
\begin{equation}
\left( p\,+  \frac{%
2\gamma _{1}}{R_{m_{1}}}\,\right)\mathbf{n}_{1} - \mathbf{\tau }_{v} \mathbf{n}_{1}+\left(\mathbf{1}-%
\mathbf{n}_{1}\mathbf{n}_{1}^T \right) \func{grad} \gamma _{1}  +\mathbf{T}_{1}=\chi\, \mathbf{n}%
_{1} ,  \label{CLI}
\end{equation}
The normal and tangential components of Eq. (\ref{CLI}) relative to $S _{1}$
are deduced from Eq. (\ref{CLI}),
\begin{eqnarray}&&
\displaystyle \ p+ \frac{%
2\gamma _{1}}{R_{m_{1}}}-\mathbf{n}_{1}^{T}\mathbf{\tau }_{v}\,\mathbf{n}_{1}+ \mathbf{n}_{1}^T\, \mathbf{T}_{1} =  \chi,
\label{CL1a} \\
&& \left( \mathbf{1}-\mathbf{n}_{1}\mathbf{n}_{1}^{T}\right) \left(- \mathbf{\tau%
}_{v}\mathbf{n}_{1}+  \func{grad} \gamma
_{1} +   \mathbf{T}%
_{1}\right)  =  0\, .  \label{CL1b}
\end{eqnarray}%
Following Eq. (\ref{CL1a}), we obtain the value of $\chi$ along the surface $%
S _{1}$. The scalar field $\chi$ corresponds to the unknown value of the
normal stress vector on the surface $S _{1}$; it corresponds to the
difference between the mechanical and viscous normal stresses and a stress
due to the curvature of $S_1$ taking account of the surface tension.
Equation (\ref{CL1b}) represents the balance between the tangential
components of the mechanical and viscous stresses and   the tangential
component of the surface tension gradient.

$\bullet \quad $ We get \textit{the condition} on surface $S _{2}$ %
\begin{equation}
\left(p\,+\frac{%
2\gamma _{2}}{R_{m_{2}}}\,\right)\mathbf{n}_{2} - \mathbf{\tau }_{v} \mathbf{n}_{2}+ \left(\mathbf{1}-%
\mathbf{n}_{2}\mathbf{n}_{2}^T \right) \func{grad} \gamma _{2}    -P\,\mathbf{n}_{2}=0.  \label{CLII}
\end{equation}%
The normal and tangential components of Eq. (\ref{CLII}) relative to $S _{2}$
are deduced
\begin{eqnarray}
\displaystyle && \frac{2\gamma _{2}}{R_{m_{2}}} - \mathbf{n}_{2}^{T}\mathbf{%
\tau }_{v}\,\mathbf{n}_{2}+ p  =  P ,  \label{CL2a} \\
 && \left( \mathbf{1}-\mathbf{n}_{2}\mathbf{n}_{2}^{T}\right) \left(-\mathbf{\tau%
}_{v}\mathbf{n}_{2}+  \func{grad} \gamma
_{2}\right) =  0  .\label{CL2b}
\end{eqnarray}
Equation (\ref{CL2a}) corresponds to the expression of the Laplace equation
in case of viscous motions; the normal component of viscous stresses is taken into
account. Equation (\ref{CL2b}) is similar to Eq. (\ref{CL1b}) for the
surface $S_2$ but without component  of the stress vector.

$\bullet \quad $ We get \textit{the condition} on line $\Gamma$

To get the line condition we must consider a virtual displacement  tangent to $S_1$ and consequently in the form
\begin{equation*}
\mathbf{\zeta}=\alpha\, \mathbf{t}+\beta\, \mathbf{t}\times \mathbf{n}_{1},
\end{equation*}
where $\alpha $\ and $\beta $\ are two scalar fields defined on $\Gamma$. From the
last integral of Eq. (\ref{works}), we get immediately: \newline
For any scalar field $\mathbf{x\in }$ $\Gamma \longrightarrow \beta \left(
\mathbf{x}\right) \in \Re ,$
\begin{equation*}
\int_{\Gamma}\beta \gamma _{1}\,\mathbf{n}_{1}^{\prime T}\left( \mathbf{t}%
\times \mathbf{n}_{1}\right) ~dl-\int_{\Gamma}\beta \gamma _{2}\,\mathbf{n}%
_{2}^{\prime T}\left( \mathbf{t}\times \mathbf{n}_{1}\right) ~dl=0,
\end{equation*}%
with $\mathbf{n^{\prime}}_1 = -\mathbf{t}\times \mathbf{n}_{1}$ and $\mathbf{n^{\prime}}_2=\mathbf{t}\times \mathbf{n}_{2}$  and consequently,
\begin{equation*}
-\gamma _{1}-\gamma _{2}\,\mathbf{n}_{2}^{T}\mathbf{n}_{1}=0.  \label{Young1}
\end{equation*}
Denoting by $\theta$ the angle $<\mathbf{n}_{1},\mathbf{n}_{2}>$, we
obtain the well-known relation of Young-Dupr\'e but adapted to $\gamma_1$ and $\gamma_2$ in
place of $\sigma_1$ and $\sigma_2$
\begin{equation}
\gamma _{1}+\gamma _{2}\cos \theta =0.  \label{Young2}
\end{equation}

\subsection{\textbf{Remarks}}

For a motionless fluid, ${%
\mathbf{\tau}}_v=0$ and consequently:

Equation (\ref{CL1b}) yields,
\begin{equation*}
\mathrm{grad}_{tg}\gamma _{1}=-\mathbf{T}_{1tg},
\end{equation*}
where $\mathrm{grad}_{tg}\gamma _{1}$ and $\mathbf{T}_{1tg}$ denote  the tangential parts of $\mathrm{grad}\, \gamma _{1}$ and $\mathbf{T}_1$, respectively. The
tangential part of the vector stress is opposite to the surface tension
gradient. Therefore, at given value of $T_{1n}= \mathbf{n}_1^T\mathbf{T}_1 $,
Eq. (\ref{CL1a}) yields the value $\chi $ corresponding to the normal stress vector to the surface $S _{1}$,\newline
 Equation (\ref{CL2a}) yields  $\displaystyle P = p +\frac{2\, \gamma_2}{R_{m_2}}$  corresponding to the classical equation of
Badshforth and Adams  \cite{Adamson} but with the  surface tension  $\gamma_2$ instead of $\sigma_2$,\newline
Equation (\ref{CL2b}) implies $\left( \mathbf{1}-\mathbf{n}_{2}\mathbf{n}%
_{2}^{T}\right)\func{grad}\gamma _{2} =0$. At equilibrium, along $S_2$, the
surface tension $\gamma _{2}$ must be uniform.

 In case of motion, Eq. (\ref%
{CL2b}) represents the Marangoni effect as proposed in \cite{Defay,Gibbs}  but with the  surface tension  $\gamma_2$ instead of $\sigma_2$.%
\newline

\section{Surface energy and surface tension}

A \emph{surface tension}   must appear on the boundary conditions as a force
per unit of line. The Legendre transformation $\gamma$ of $\sigma$ with
respect to $c$ exactly corresponds  to this property on the contact line $\Gamma$; then, \emph{surface tension} $%
\gamma$   differs from the \emph{surface energy}; this important
property was pointed out by Gibbs \cite{Gibbs} and Defay \cite{Defay} by
means of thermodynamical considerations. The fundamental difference between
surface tension and surface energy, in presence of attached surfactants, is
illustrated in the following cases corresponding to formal behaviors. %
\newline
-\quad If $\sigma $ is independant of $c$, then $\gamma =\sigma $\thinspace : the
surface tension is equal to the surface energy. This is the classical
case of capillarity for fluids considered in Section 2.2  and Eq. (\ref%
{Young2}) is the classical Young-Dupr\'e condition on the contact lines.%
\newline
-\quad In fact $\sigma$ is a decreasing function of $c$ \cite{Adamson}; when $c$ is
small enough we consider the behavior
\begin{equation*}
\sigma =\sigma _{0}-\sigma _{1}\,c \quad \mathrm{where}\ \sigma _{0} > 0 \
\mathrm{and}\, \ \sigma _{1} > 0 \, ,
\end{equation*}
then, Eq. (\ref{legendre}) implies $\gamma =\sigma _{0}$ and surface tension and surface energy are
different.\newline
-\quad Now, we consider a formal case when the surface energy density model writes in
the form
\begin{equation*}
\sigma =\sigma _{0}-\sigma _{1}\,c -\sigma _{2}\,c\,\sin \left( \frac{1}{c}%
\right)
\end{equation*}
where $\sigma _{0} > 0,\, \sigma _{1}> 0,\, \sigma _{2}> 0$.
Then, Eq. (\ref{legendre}) implies
\begin{equation}
\gamma =\sigma _{0}-\sigma _{2}\cos \left( \frac{1}{c}\right).
\label{hysteresis}
\end{equation}
This case does not
correspond to $\sigma$ as a monotonic decreasing function of $c$. Nevertheless,
when $c$ $\rightarrow 0$, $\gamma $ does not have any limit and we get
\begin{equation*}
\gamma \in \left[ \sigma _{0}-\sigma _{2},\sigma _{0}+\sigma _{2}\right] .
\end{equation*}
The surface tension may have a large scale of values. When the concentration
$c$ is low, a variation of the concentration $c$ may generate strong
fluctuations of the surface tension without significant change of the
surface energy. Alternatively,   the concentration behavior
strongly affects the surface tension but not the surface energy.
Relation (\ref{hysteresis}) fits with the well-known physical case of an hysteresis behavior for a drop
lying on a horizontal plane
(see for example \cite{Gouin2} and the literature therein). So, \emph{the surface
roughness is not the only reason of the hysteresis of the contact angle even if the surface energy is nearly constant.}

\section{Conclusion}

The principle of virtual work allows us to deduce the equation of motion and
conditions on surfaces and line by means of a variational analysis. When
capillary forces operate and surfactant molecules are
attached to the fluid molecules at the interfaces, the conditions on
surfaces and lines point out a fundamental difference between the concepts
of surface energy and surface tension. This fact was thermodynamically
predicted in \cite{Defay,Gibbs}. Hysteresis phenomenon may appear even if
surface energy is almost constant on a planar substrate when the surface
tension strongly varies.\newline
In Eq. (\ref{Young2}), $\gamma _{1}$ and $\gamma _{2}$
are not assumed to be constant, but are defined at each point of $%
\Gamma $. This expression of Young-Dupr\'e boundary condition on the contact line $%
\Gamma$ is not true in more complex cases. For example in the case when the
surface tension is a non-local functional of surfactant concentration, the
surface tension is no longer the classical Legendre transformation of the
surface energy relative to surfactant concentration and more complex
behaviors can be foreseen. These behaviors can change the variation of the
integral of the free energy as in the case of shells or in second gradient
models for which boundary conditions become more complex \cite%
{Germain1,Cosserat,Toupin,Noll,Isola}. In a further article \cite{Gouin 8}, we will see  this is also the case when the
surface energy depends on the surface curvature as in membranes and
vesicles \cite{Helfrich,Seifert,Steigmann2}.
\\

\section{{Appendix} - Geometrical preliminaries \protect\cite%
{Kosinski,Aris,Koba}}

\subsection{\textbf{Expression of the virtual work of forces in capillarity}}

The hypotheses and notations are presented in the previous Section 2.2.

\subsubsection{\textbf{Lemma 1}: we have the following relations,}

\begin{eqnarray}
\delta \det F &=&\ \det F\, \func{div}\mathbf{\zeta }\,,  \label{Jacobi} \\
\delta \left( F^{-1}\mathbf{n}\right) &=&-F^{-1}\ \frac{\partial \mathbf{%
\zeta }}{\partial \mathbf{x}}\,\mathbf{n\ +\ }F^{-1}\, \delta\mathbf{n}\,.  \label{InverseF}
\end{eqnarray}%
The proof of Rel. (\ref{Jacobi}) comes from the  so-called Jacobi identity
\begin{equation*}
\delta (\det F)=\det F\,\mathtt{tr}\left( F^{-1}\delta F\right)
\end{equation*}%
and from
\begin{equation*}
\delta F=\delta \left( \frac{\partial \mathbf{x}}{\partial \mathbf{X}}%
\right) =  \frac{\partial \mathbf{\zeta }}{\partial \mathbf{X}} .
\end{equation*}%
Then,
\begin{equation*}
\mathtt{tr}\left( F^{-1}\delta F\right) =\mathtt{tr}\left( \frac{\partial
\mathbf{X}}{\partial \mathbf{x}}\,\frac{\partial \mathbf{\zeta }}{\partial
\mathbf{X}}\right) =\mathtt{tr}\left( \frac{\partial \mathbf{\zeta }}{%
\partial \mathbf{X}}\,\frac{\partial \mathbf{X}}{\partial \mathbf{x}}\right)
=\mathtt{tr}\left( \frac{\partial \mathbf{\zeta }}{\partial \mathbf{x}}%
\right) =\mathrm{div}\,{\mathbf{\zeta }}.
\end{equation*}%
The proof of Rel. (\ref{InverseF}) comes as follow
\begin{equation*}
\delta \left( F^{-1}\mathbf{n}\right) =\delta \left( F^{-1}\right) \mathbf{n}%
+F^{-1}\delta \mathbf{n}
\end{equation*}%
But the implication,
\begin{equation*}
F^{-1}\,F=\mathbf{1}\ \Longrightarrow \ \delta \left( F^{-1}\right)
F+F^{-1}\,\delta F=0\ \Longrightarrow \ \delta \left( F^{-1}\right) =-F^{-1}%
\frac{\partial \mathbf{\zeta }}{\partial \mathbf{X}}F^{-1}= - F^{-1}\frac{%
\partial \mathbf{\zeta }}{\partial \mathbf{x}},
\end{equation*}%
yields Rel. (\ref{InverseF}).

\subsubsection{{\textbf{Lemma 2} }: {Let us consider the surface integral} $
\ \displaystyle E=\protect\int \protect\int_{S}\protect\sigma\,ds.$ {Then the
variation of} $E$ {is},}
\begin{equation}
\delta E=\int \int_{S}\left[ \delta \sigma-\left( \frac{2\sigma%
}{R_{m}}\,\mathbf{n}^{T}+\mathrm{{grad}^{T}\sigma\left( \mathbf{1}-\mathbf{nn}^{T}\right) }%
\right) \mathbf{\zeta}\right] ~ds+\int_{\Gamma }\sigma\,\mathbf{n}^{\prime T}%
\mathbf{\zeta}~dl .  \label{varsurf0}
\end{equation}

Relation (\ref{varsurf0}) points out the extreme importance to know the
variation of $\delta \sigma$. The variation $\delta E$ of $E$ drastically
changes following the different possible behaviors of the surface energy.

The proof can be found as follows: the external normal ${\mathbf{n(x)}}$ to $S$ is locally extended in the
vicinity of $S$ by the relation ${\mathbf{n(x)}} = \mathrm{grad}\ d(\mathbf{x%
}), $ where $d$ is the distance of  point $\mathbf{x}$ to $S$; for any
vector field ${\mathbf{w}}$, we obtain \cite{Aris,Koba}\,
\begin{equation*}
\mathrm{rot} ({\mathbf{n}} \times {\mathbf{w}}) ={\mathbf{n }}\, \mathrm{div}
\, {\mathbf{w}}- {\mathbf{w}}\,\mathrm{div}\,{\mathbf{n }} + \frac {\partial
{\mathbf{n}}} {\partial {\mathbf{x}}}\, {\mathbf{w}}- \frac {\partial {%
\mathbf{w}}} {\partial {\mathbf{x}}}\, {\mathbf{n}}.
\end{equation*}
From $\,\displaystyle {\mathbf{n}}^T\frac {\partial {\mathbf{n}}} {\partial {%
\mathbf{x}}}= 0\,$ and $\,\mathrm{div}\,{\mathbf{n }} =\displaystyle -\frac {%
2} {R_m}$\,,\ we deduce on $S$,
\begin{equation}
{\mathbf{n}^T} \mathrm{rot} ({\mathbf{n}} \times {\mathbf{w}}) = \mathrm{div}
\, {\mathbf{w}}+\frac {2} {R_m}\, {\mathbf{n}^T} {\mathbf{w}} - {\mathbf{n}^T%
} \frac {\partial {\mathbf{w}}} {\partial {\mathbf{x}}}\, {\mathbf{n}}.
\label{A0}
\end{equation}
Due to $\displaystyle\ E=\int\int_S \sigma\ \det\, ({\mathbf{n}},d_1{\mathbf{%
x}},d_2{\mathbf{x}})\ $ where $\displaystyle\ d_1{\mathbf{x}} \ $ and $%
\displaystyle\ d_2{\mathbf{x}}\ $ are differentiable vectors associated with two  coordinate lines of $S, $ we get
\begin{equation*}
E =\int\int_{S_0}\sigma\, \det\, F\ \hbox{det}\, (F^{-1}{\mathbf{n}},d_1{%
\mathbf{X}},d_2{\mathbf{X}}) ,
\end{equation*}
where $d_1{\mathbf{x}}=F\,d_1{\mathbf{X}}$ and  $d_2{\mathbf{x}}=F\,d_2{\mathbf{X}}$. Then,
\begin{equation*}
\displaystyle \delta E =\int\int_{S_0}\delta \sigma\ \det\, F\ \hbox{det}\,
(F^{-1}{\mathbf{n}},d_1{\mathbf{X}},d_2{\mathbf{X}}) +
\int\int_{S_0}\sigma\, \delta \left(\det\, F\ \hbox{det}\, (F^{-1}{\mathbf{n}%
},d_1{\mathbf{X}},d_2{\mathbf{X}})\right ).
\end{equation*}

Due to Lemma 1 and to $\,\displaystyle {\mathbf{n}}^T \mathbf{n}= 1\, \Longrightarrow\, {\mathbf{n}}^T \delta\mathbf{n}= 0$,
\begin{equation*}
\begin{array}{cc}
\displaystyle \int\int_{S_0}\sigma\, \delta \left(\det\, F\ \hbox{det}\,
(F^{-1}{\mathbf{n}},d_1{\mathbf{X}},d_2{\mathbf{X}})\right) = &  \\
\displaystyle \int\int_S \left[\sigma\ \mathrm{div}\, {\mathbf{\zeta}} \ \det ({%
\mathbf{n}},d_1{\mathbf{x}},d_2{\mathbf{x}}) + \sigma\, \det \left (%
\displaystyle \delta{\mathbf{n}},d_1{\mathbf{x}},d_2{\mathbf{x}}\right ) \displaystyle -
\sigma \det \left (\displaystyle \frac {\partial{\mathbf{\zeta}}}{\partial {%
\mathbf{x}}}\,{\mathbf{n}},d_1{\mathbf{x}},d_2{\mathbf{x}} \right ) \right]= &  \\
\displaystyle \int\int_S \left( \mathrm{div} (\sigma\,{\mathbf{\zeta}} )-(%
\mathrm{grad}^T \sigma) \, {\mathbf{\zeta}} -\sigma\, {\mathbf{n}}^T \frac {%
\partial {\mathbf{\zeta}}} {\partial {\mathbf{x}}} \, {\mathbf{n}} \right )
ds. &
\end{array}%
\end{equation*}
Relation (\ref{A0}) yields
\begin{equation*}
\displaystyle \mathrm{div}\, (\sigma\,{\mathbf{\zeta}}) + \frac {2\sigma} {%
R_m}\, {\mathbf{n}}^T {\mathbf{\zeta}} - {\mathbf{n}}^T \frac {\partial
\sigma  {\mathbf{\zeta}}} {\partial {\mathbf{x}}} \, {\mathbf{n}} = {%
\mathbf{n}}^T\, \mathrm{rot}\, (\sigma\,{\mathbf{n}}\times {\mathbf{\zeta}}
) .
\end{equation*}
Then,
\begin{equation*}
\begin{array}{cc}
\displaystyle \int\int_{S_0}\sigma\,\delta \left (\det\ F\,\hbox{det}\,
(F^{-1}{\mathbf{n}},d_1{\mathbf{X}},d_2{\mathbf{X}})\right ) = &  \\
\displaystyle\ \int\int_{S} \left( - \frac {2\sigma} {R_m} \, {\mathbf{n}}^T+%
\mathrm{grad}^T\sigma\,({\mathbf{nn}^T-\mathbf{1}}) \right ) {\mathbf{\zeta}}
\, ds+\int\int_S{\mathbf{n}}^T\ \mathrm{rot}\, (\sigma\,{\mathbf{n}}\times{%
\mathbf{\zeta}})\,ds, &
\end{array}%
\end{equation*}

where $\mathrm{grad}^{T}\sigma \,({{\mathbf{nn}}^{T}-\mathbf{1}}) $ belongs to the cotangent plane to $S$ and
we obtain Relation (\ref{varsurf0}).

\subsubsection{\textbf{Variation of the internal energy}}

Let us note that\ $\displaystyle\delta\int \int \int_{D}\rho \ \alpha
\,dv=\int \int \int_{D}\rho \ \delta \alpha \,dv$\ where $\displaystyle\
\delta \alpha =\frac{\partial \alpha }{\partial \rho }\ \delta \rho $.
\newline
Due to the mass conservation,
\begin{equation}
\rho \ \mathrm{det}\,F=\rho _{0}(\mathbf{X}),  \label{mass}
\end{equation}%
where $\rho _{0}$ is defined on $D_{0}$, the differentiation of Eq. (\ref%
{mass}) yields,
\begin{equation*}
\delta \rho \,\mathrm{det}F+\rho \,\delta (\mathrm{det}F)=0
\end{equation*}%
and from Lemma 1, we get
\begin{equation*}
\delta \,\rho =-\rho \,\mathrm{div}\,{\mathbf{\zeta }}.
\end{equation*}%
Consequently, from $\displaystyle p = \rho^2 \frac{\partial \alpha }{\partial \rho }$ and $\mathrm{div} (p\, {\mathbf{\zeta }}) = p \, \mathrm{div}   {\mathbf{\zeta }} + (\mathrm{grad}\,p)^{T}\,{\mathbf{\zeta }}$ ,\ we get
\begin{eqnarray}
\delta \int \int \int_{D}\rho \ \alpha \,dv &=&  \label{pressure} \\
\int \int \int_{D}\rho \, \frac{\partial \alpha }{\partial \rho }\,\delta
\rho \,dv &=&\int \int \int_{D}-p\ \mathrm{div}\,{\mathbf{\zeta }}\,dv=\int
\int \int_{D}(\mathrm{grad}\,p)^{T}\,{\mathbf{\zeta }}\,dv-\int \int_{S}{p\,%
\mathbf{n}}^{T}{\mathbf{\zeta }}\,ds.  \notag
\end{eqnarray}%
By taking account of Rel. (\ref{varsurf0}), we immediately get Rel. (\ref%
{Laplace}).

\subsection{\textbf{Study of a surfactant attached to fluid particles}}
\subsubsection{ \textbf{Proof of relation}  (\ref{concentration2})}
Under the hypotheses and notations of Section 3.2,
\begin{eqnarray*}
\int \int_{S^{\ast }}c\ ds &=& \int \int_{S^{\ast }} {\det} ({\mathbf{n}}\,
c, d_1{\mathbf{x}}, d_2{\mathbf{x}}) =\int \int_{S^{\ast }_0} {\det} ({F
F^{-1}\mathbf{n}}\, c, F d_1{\mathbf{X}},F d_2{\mathbf{X}}) \\
&=& \int \int_{S_{0}^{\ast }} c\, ({\det}F)\, \mathrm{det} ({\ F^{-1}\mathbf{%
n}}, d_1{\mathbf{X}}, d_2{\mathbf{X}}) = \int \int_{S_{0}^{\ast }}c\, (\det
F)\ \mathbf{n}_{0}^{T}F^{-1}\mathbf{n}~ds_{0},
\end{eqnarray*}
where $\mathbf{n}_{0}^{T}\, \mathbf{n}_{0} = 1$. Moreover, $\mathbf{n}^{T} d{%
\mathbf{x}} =0 \Rightarrow \mathbf{n}^{T} F d{\mathbf{X}} =0$, then $\mathbf{%
n^{\prime }}_0^{T} = \mathbf{n}^{T} F$ is normal to $S_{0}^{\ast }$ and
consequently,
\begin{equation*}
\mathbf{n}_{0}^{T}=\frac{\mathbf{n}^{T}F}{\sqrt{(\mathbf{n}^{T}FF^{T}\mathbf{%
n})}} \,, \qquad \mathbf{n}^T =\frac{\mathbf{n}_0^T F^{-1}}{\sqrt{(\mathbf{n}^{T}_0F^{-1}({F^{-1}})^T \mathbf{n_0})}}
\end{equation*}
and from Rel. (\ref{surfactint}),
\begin{equation}
c\,  \det F \ \sqrt{\mathbf{n}_{0}^{T}F^{-1}({F^{-1}})^T \mathbf{n_0}}=c_{0}.\label{co}
\end{equation}

\subsubsection{\textbf{Proof of relations} (\protect\ref{conservation}) and (%
\protect\ref{varsurfactant})}

With the notations of Section 3.2, Rel. (\ref{co}) yields
\begin{equation*}
\frac{dc}{dt}= \displaystyle - \frac{c_0\,\displaystyle\frac{d\,(\det F)}{dt}
}{\left(\det F\right)^2 \sqrt{\big(\mathbf{n}_0^{T}F^{-1}(F^{-1})^T \mathbf{%
n }_0\big)}}- \frac{c_0\,\displaystyle\frac{d}{dt}\left(\mathbf{n}%
_0^{T}F^{-1}(F^{-1})^T\mathbf{n }_0\right)}{2 \det F \big(\mathbf{n}%
_0^{T}F^{-1}(F^{-1})^T \mathbf{n }_0\big)^{{3}/{2}}} .
\end{equation*}
But, $\displaystyle\frac{d\,(\det F)}{dt}= (\det F)\, \mathrm{div}\, \mathbf{u}
$ \ and \ $\displaystyle\frac{d}{dt}\left(F^{-1}(F^{-1})^T\right) = - 2
F^{-1} \mathbf{D}\, (F^{-1})^T$. Then,
\begin{equation*}
\frac{dc}{dt}+c\left( \func{div}\mathbf{u}-\mathbf{n}^{T}\mathbf{D}\ \mathbf{%
n}\right) =0.
\end{equation*}
The same calculation with $\delta$ in place of $\ \displaystyle\frac{d}{dt}\ $
yields immediately
\begin{equation*}
\delta c+c\left[ \func{div}\mathbf{\zeta}-\mathbf{n}^{T}\frac{\partial
\mathbf{\zeta}}{\partial \mathbf{x}}\,\mathbf{n}\right] =0.
\end{equation*}

\subsubsection{\textbf{Proof of relation} (\protect\ref{variationE})}
From Rel. (\ref{varsurfactant}) and $\sigma = \sigma (c)$ we get,
\begin{equation*}
\delta \sigma=\kappa\left[ \func{div}\mathbf{\zeta}-\mathbf{n}^{T}\frac{\partial
\mathbf{\zeta}}{\partial \mathbf{x}}\,\mathbf{n}\right]\quad {\rm with} \quad \kappa(c) = -c\, \sigma^{\prime }(c).
\end{equation*}
Consequently,
\begin{equation*}
\int\int_S \delta\sigma= \int\int_S \left(\func{div}(\kappa\,\mathbf{\zeta})- \mathrm{grad}^T\kappa \ \mathbf{\zeta}-\kappa\,\mathbf{n}^{T}\frac{\partial
\mathbf{\zeta}}{\partial \mathbf{x}}\,\mathbf{n}\right)ds.
\end{equation*}
But Rel. (\ref{A0})  implies,
\begin{equation*}
{\mathbf{n}^T} \mathrm{rot} (\kappa\, {\mathbf{n}} \times {\mathbf{\zeta}}) = \mathrm{div}(
\kappa\, {\mathbf{\zeta}})+\frac {2\kappa} {R_m}\, {\mathbf{n}^T} {\mathbf{\zeta}} - {\mathbf{n}^T%
} \frac {\partial {(\kappa\,\mathbf{\zeta})}} {\partial {\mathbf{x}}}\, {\mathbf{n}}
\end{equation*}
and,
\begin{equation*}
{\mathbf{n}^T%
} \frac {\partial {(\kappa\,\mathbf{\zeta})}} {\partial {\mathbf{x}}}\, {\mathbf{n}}= (\mathbf{n}^T\mathbf{\zeta}) \textbf{.} ( \mathrm{grad}^T\kappa \ \mathbf{n})+ \kappa\,\mathbf{n}^{T}\frac{\partial
\mathbf{\zeta}}{\partial \mathbf{x}}\,\mathbf{n}= \mathrm{grad}^T\kappa \, \mathbf{n}  \mathbf{n}^{T} \mathbf{\zeta}+ \kappa\,\mathbf{n}^{T}\frac{\partial
\mathbf{\zeta}}{\partial \mathbf{x}}\,\mathbf{n}.
\end{equation*}
Then,
\begin{equation*}
\func{div}( \kappa\,\mathbf{\zeta})- \mathrm{grad}^T\kappa \ \mathbf{\zeta}-\kappa\,\mathbf{n}^{T}\frac{\partial
\mathbf{\zeta}}{\partial \mathbf{x}}\,\mathbf{n}=-\frac {2\kappa} {R_m}\, {\mathbf{n}^T} {\mathbf{\zeta}}-\mathrm{grad}^T\kappa \,(\textbf{1}- {\mathbf{n}}{\mathbf{n}^T}) \mathbf{\zeta}+{\mathbf{n}^T} \mathrm{rot} (\kappa\, {\mathbf{n}} \times {\mathbf{\zeta}})
\end{equation*}
Due to
\begin{equation*}
\int\int_S {\mathbf{n}^T} \mathrm{rot} (\kappa\, {\mathbf{n}} \times {\mathbf{\zeta}})\,ds = \int_{\Gamma }\kappa\,\mathbf{n}^{\prime T}%
\mathbf{\zeta}~dl,
\end{equation*}
we get
\begin{equation*}
\int\int_S -\delta\sigma\,ds=\int\int_S\left[\,\frac {2\kappa} {R_m}\, {\mathbf{n}^T} +\mathrm{grad}^T\kappa \,(\textbf{1}- {\mathbf{n}}{\mathbf{n}^T})  \right]\mathbf{\zeta}\,ds- \int_{\Gamma }\kappa\,\mathbf{n}^{\prime T}%
\mathbf{\zeta}~dl,
\end{equation*}
and Rel. (\ref{varsurf0}) yields,
\begin{equation*}
\delta E=-\int \int_{S}\left[ \, \frac{2\gamma }{R_{m}}\,\mathbf{n}^{T}+%
\mathrm{{grad}^{T}\gamma \ \left( \mathbf{1}-\mathbf{nn}^{T}\right) }
\right]\mathbf{\zeta} ~ds+\int_{\Gamma}\gamma \ \mathbf{n}^{\prime T}\mathbf{%
\zeta}~dl.
\end{equation*}

\end{document}